# Decoding local framework dynamics in the ultra-small pore MOF MIL-120(Al) CO$_2$ sorbent with Machine Learned Potentials

Dong Fan,[1,2] Felipe Lopes Oliveira,[2] Mohammad Wahiduzzaman,[2] Guillaume Maurin[2,3,*]

[1] School of Materials Science and Engineering, Chongqing Jiaotong University, Chongqing 400074, P.R. China

[2] ICGM, Univ. Montpellier, CNRS, ENSCM, Montpellier, 34095, France

[3] Institut Universitaire de France, France.

*Corresponding author: guillaume.maurin1@umontpellier.fr

## Abstract

Metal–organic frameworks (MOFs) with ultra-small pores offer an optimal environment to effectively capture guest molecules such as CO$_2$. Subtle local dynamics of their frameworks, either throughout reorientation of functional groups grafted to the organic linkers or those present in their inorganic nodes, is expected to play a major role in their sorption behaviors. Here, we combine density-functional theory (DFT) with a purpose-trained machine-learned potential to systematically investigate the local dynamics of the bridging hydroxyl groups, $\mu_2$-OH groups present in the prototypical ultra-small pore MOF MIL-120(Al), reported recently as an attractive CO$_2$ sorbent. We identified six MOF configurations associated with distinct $\mu_2$-OH orientations with relatively low interconversion energy barriers (0.07–0.19 eV per unit cell) suggesting that all these states can be observed experimentally at room temperature. We demonstrated that our MLP achieves near-DFT-level fidelity, reproducing the energy barriers and phonon spectra of the empty MOF, and accurately predicting CO$_2$ adsorption geometries depending on the $\mu_2$-OH orientations with CO$_2$ adopting either parallel or perpendicular alignment to the pore axis, which in turn governs the adsorption energetics. This work establishes that a reliable description of the local structure, such as reorientation/flipping of bridging hydroxyl groups, is a key feature to gain an accurate description of the guest locations and energetics in ultra-small pore MOFs.

**Introduction**

Metal–organic frameworks (MOFs) are crystalline coordination polymers constructed from metal ions or clusters connected by organic linkers, forming highly ordered nanoporous structures.[1,2] Owing to the unique tunability of their pore size/shape/chemical functionality, MOFs have emerged as promising materials for diverse applications including $CO_2$ capture.[3,4] In particular, MOFs with ultra-small pores/channels combined potentially with polar groups decorating the pore walls offer a unique confined environment to favour an effective packing of $CO_2$ in the pores. This confers to this sub-class of MOFs attractive $CO_2$ sorption performance even in the presence of co-adsorbed species such as $N_2$ or $CH_4$ of key importance in the context of $CO_2$ capture in post- and pre-combustion conditions.[5–8] Notably, it has been documented that a very tiny change of the pore size of isoreticular ultra-small pore MOFs by modulating the nature of the metal sites as for example in the KAUST-8 series (Al, Fe, Ga) can fine-tune their $CO_2$ sorption properties.[9,10] Local flexibility of such a sub-class of MOFs that can arise either from the dynamics of the functional groups grafted to the organic linkers or the orientation of the chemical functions present in the inorganic nodes can also play a key role in the $CO_2$ sorption mechanism.[11–14] This structural dynamics is often overlooked, although decisive in controlling the $CO_2$/MOF interactions by modulating slightly the MOF pore confinement. This is especially true for ultra-small pore MOFs containing bridging hydroxyl groups, namely, $\mu_2$-OH groups, with a notable experimental limitation lying in determining hydrogen atom positions since H atoms scatter X-rays only weakly and their locations cannot be detected by X-ray diffraction. As a result, hydrogen atoms are typically added post hoc based on standard bond lengths and geometries.[15] Such empirical treatments assume that $\mu_2$-OH orientations exert only a minor influence on MOF framework properties, which might be only valid for medium- to large-pore MOFs. MIL-120(Al), first reported in 2009 by Férey and co-workers, is one representative ultra-small pore MOF.[16] This aluminum MOF is built from the assembly of tetratopic linker 1,2,4,5-benzenetetracarboxylate and infinite chains of edge-sharing $AlO_6$ octahedra[16] forming a three-dimensional network with one-dimensional ultra-small pores (~5.4 × 4.7 Å) aligned along the c-axis. Combined with its low-cost hydrothermal synthesis and moderate surface area, this MOF was demonstrated recently to exhibit attractive $CO_2$ adsorption performance, maintaining capture capacities of ~1.2 mmol g$^{-1}$ even under humid flue-gas conditions.[16–18] This combination of hydrolytic stability, ultra-microporosity, and strong $CO_2$ affinity has established MIL-120(Al) as one of the most promising Al-MOFs for cost-effective carbon capture.[18] Our preliminary findings suggested that the orientation of the $\mu_2$-OH groups present in this MOF affects the $CO_2$ location in the pores and their associated energetics. Nevertheless, a systematic exploration of the role played by the local dynamics of these $\mu_2$-OH groups on the $CO_2$ sorption properties of this MOF, is yet to be realized.

To address this gap, the present work employs a combination of density-functional theory (DFT) calculations and a machine-learned potential (MLP) approach to systematically map the configurational landscape of $\mu_2$-OH groups in MIL-120(Al). By quantifying their relative electronic stabilities, transition pathways, and impact on $CO_2$ adsorption thermodynamics, this study not only reveals a previously overlooked structural degree of freedom in MIL-120(Al), but also establishes a transferable computational framework for probing the role of functional group orientations in MOFs more broadly.

**Results**

***Exploration of the local structural features of MIL-120(Al).*** The structure of MIL-120(Al) comprises edge-sharing $AlO_6$ octahedra linked by bridging hydroxyl ($\mu_2$-OH) groups forming one-dimensional chain motifs (*cf*. Fig. 1a). Among several possible $\mu_2$-OH groups orientations, we constructed six representative structure models of MIL-120(Al) that differ exclusively in the orientation of the four $\mu_2$-OH groups present in the unit cell, namely MIL-120(Al)-A, MIL-120(Al)-B, MIL-120(Al)-C, MIL-120(Al)-D, MIL-120(Al)-E and MIL-120(Al)-F (*cf.* Figs. 1b) while the MOF skeleton remains the same as shown in Fig. 1c with identical simulated X-ray diffraction patterns for the six constructed structure models in line with the corresponding experimental data.[17] Notably, because X-ray diffraction cannot resolve the positions of hydrogen atoms, previous studies have adopted MIL-120(Al)-F as a single representative structure model.[16,18] The DFT-geometry optimization with a full relaxation of the atomic positions and cell parameters of these six structure models revealed a clear distinction in terms of their total electronic energies: MIL-120(Al)-A is the most energetically favorable structure, while MIL-120(Al)-F shows the highest total electronic energy. These two structures differ in energy by approximately 0.59 eV per unit cell, indicating that the MIL-120(Al)-A configuration is significantly more stable and this remain valid even at room temperature. Interestingly, the MIL-120(Al)-F configuration widely referred to in the literature[16,18] for this MOF appears to correspond to an orientation of $\mu_2$-OH groups not energetically favourable. In the lowest-energy ordering of MIL-120(Al)-A, the $\mu_2$-OH groups form an interlocking hydrogen-bond network between adjacent $Al(OH)_4O_2$ octahedral chains, as shown in Supplementary Fig. 1. The $\mu_2$-OH groups orientate towards the neighbouring $Al(OH)_4O_2$ chain on one side, establishing directional hydrogen bonds, while on the opposite side, they are pointing towards the channel. This cooperatively interlocking motif stabilises the configuration of MIL-120(Al)-A. In contrast, the other five variants lack such motif, with more disordered $\mu_2$-OH orientations that result in weaker cooperative hydrogen-bond contacts. These six structure models exhibit only minor variations in their lattice parameters: the relative changes in *a*, *b*, *c*, *α*, *β* and *γ* remain within ±1.4%, and the cell volumes within ±3.5%, as shown in Fig. 1d. Despite these similar crystallographic features, their pore size distributions (PSDs) deviate with the main peak distributed between ~3.8 and 4.4 Å, depending on the orientations of the $\mu_2$-OH groups towards the MOF channel (*cf.* Fig. 1e). Given the ultra-small pore nature of the 1D MOF channel, such sub-angstrom variations in the pore aperture can considerably affect guest accessibility, as well as MOF/guest interactions. Therefore, a key insight is that $\mu_2$-OH orientation represents a 'hidden' structural degree of freedom not detectable by standard X-ray diffraction techniques yet capable of influencing functional pore dimensions and energetics.

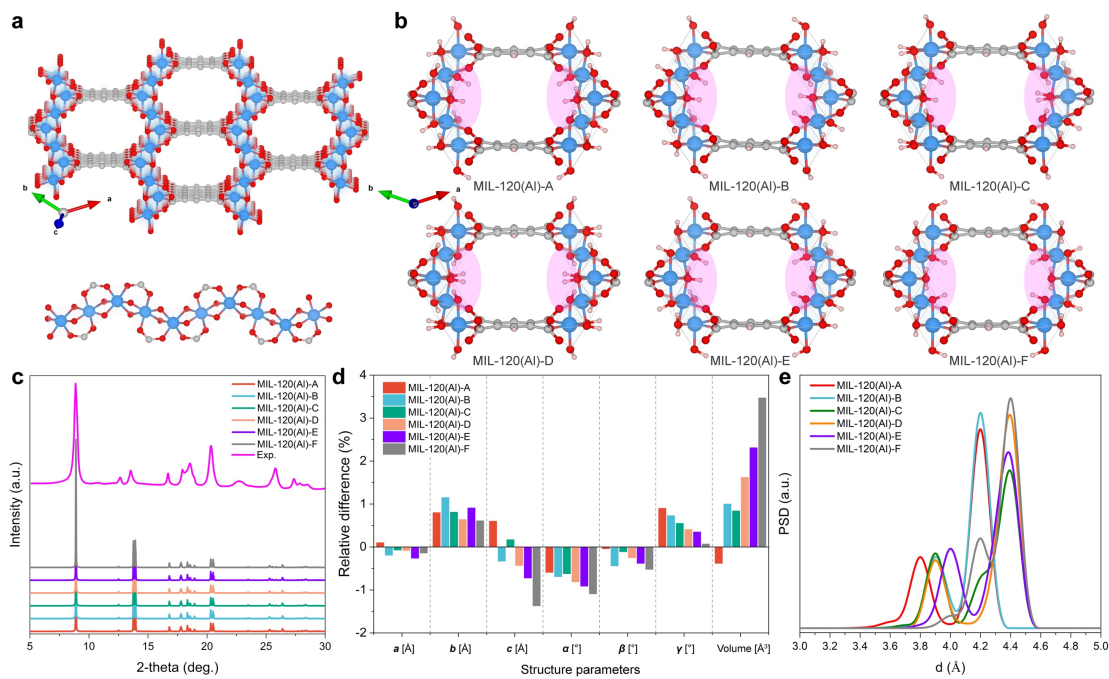

**Fig. 1 | The rich phase landscape of the MIL-120(Al) structure.** (**a**) The crystal structure of MIL-120(Al) viewed along [001] direction highlighting the ultra-small 1D channel (top). Inorganic building blocks of the *trans-cis* edge-sharing Al(OH)$_4$O$_2$ octahedra along the direction of the MOF channel (bottom). H atoms were excluded for clarity. (**b**) Representation of the DFT-optimized structure models for the six variants of MIL-120(Al) with the different orientations of the $\mu_2$-OH groups pointing towards the channel. The shadow area highlights the differences between the variants. Colour code used in the structure models: Al, blue; O, red; C, grey; H, pink. (**c**) A comparison of the X-ray diffraction patterns for the DFT-optimized MIL-120(Al) configurations and the corresponding experimental data. (**d**) The percentage difference observed in the lattice parameters for the six MIL-120(Al) structures compared to the experimental values.[17] (**e**) Computed PSDs for different MIL-120(Al)s.

*Machine-learned potential development and exploration of the structural stability of the different MIL-120(Al) configurations.* To efficiently probe the full configurational ensemble and to enable in-depth exploration of the possible transition pathways

between all these structure models, we developed a robust MLP, trained on a comprehensive MIL-120(Al)-specific DFT dataset, as shown in Supplementary Fig. 2. The dataset includes a wide range of DFT-optimised MOF structures, static single-point DFT calculations, MOF structures loaded with different $CO_2$ uptakes, transition structures obtained *via* climbing-image nudged elastic band (CI-NEB),[19] and *Ab Initio* Molecular Dynamics (AIMD) snapshots under diverse NVT/NPT conditions (details of the collection, training, validation, and testing of the entire dataset can be found in Supplementary Note1). We then trained a DeePMD potential[20] and evaluated its coverage with the t-distributed stochastic neighbor embedding (t-SNE) method. As shown in Fig. 2a, distinct datasets occupy different regions of the descriptor space, allowing for a clear view of the occupation of each component in the entire dataset. This plot enables to confirm that there is an effective sampling of the configuration space throughout the MLP training. Fig. 2b shows a comparison between the MLP-and DFT-derived energies for the empty and $CO_2$-loaded (one $CO_2$ per unit-cell, ~ 2.02 mmol g$^{-1}$) MIL-120(Al) structures with root mean square error (RMSE) values of 0.217 meV/atom and 0.268 meV/atom, respectively. Notably, these deviations are significantly lower than the values generally obtained for a standard MLP,[21,22] demonstrating the high accuracy and effectiveness of the MLP training process. The energy–volume curves derived from DFT optimization, MLP single-point evaluation based on DFT-optimized geometries, and fully MLP-relaxed structures show very good agreement, as shown in Fig. 2c in terms of both minima and overall curvatures of the plot. Phonon spectra predicted by using the MLP also show no imaginary modes at the whole Brillouin-zone path for all MIL-120(Al) variations, confirming the dynamical stability of all these configurations (*cf.* Fig. 2d). These results demonstrate that the trained MLP achieves near-DFT-level accuracy in describing the energetics, vibrational properties, and optimized geometries across the MIL-120(Al) configurations.

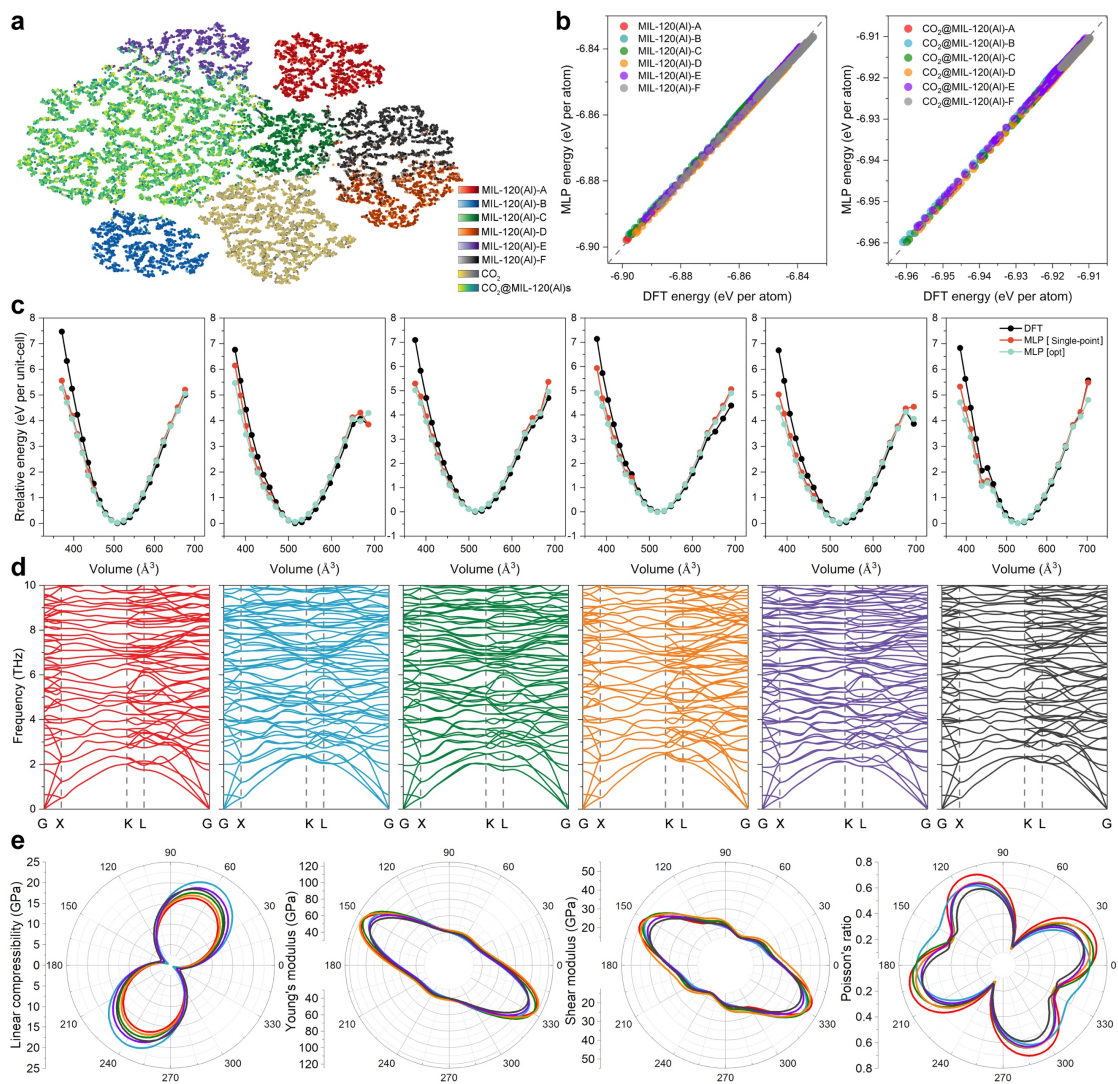

**Fig. 2 | Structural stability and mechanical properties of MIL-120(Al) configurations predicted by MLP against DFT calculations.** (**a**) t-SNE embedding of the training dataset shows distinct sampling of the $CO_2$ molecules, MIL-120(Al) structure variants, and corresponding adsorbed states. Points were coloured according to their corresponding structural types. (**b**) Linear relationship plots of MLP against DFT energies for the empty MIL-120(Al) (left) and $CO_2$-loaded MIL-120(Al) configurations (right). (**c**) Energy–volume curves computed by DFT and MLP (including single-point calculations and MLP-based relaxation) calculations for the six different MIL-120(Al) configurations. (**d**) MLP-predicted phonon spectra of the different MIL-120(Al) configurations. MIL-120(Al)-A to MIL-120(Al)-F are arranged

from left to right in sequence. (**e**) Polar coordinate diagrams of the mechanical properties of the different MIL-120(Al) configurations derived from elastic constants (along the *ab* plane). The averaged mechanical properties of all MIL-120(Al) configurations can be found in Supplementary Table 2.

We then evaluated the mechanical properties of all MIL-120(Al) configurations by determining their linear compressibility, Young's modulus, shear modulus, and Poisson's ratio. Similar to other MOFs reported earlier, such as MIL-53(Al) and CALF-20(Zn),[23,24] pronounced anisotropy of the mechanical properties was also evident for the MIL-120(Al) series (Fig. 2e). Notably, the maximum-to-minimum compressibility ratio for MIL-120(Al)-B reaches 3.6, indicating substantial structural heterogeneity. Directionally resolved stress-strain calculations confirm this anisotropic elasticity and failure behaviours. For example, the [001] orientation shows exceptional ductility and toughness, with all MIL-120(Al) configurations demonstrating ultimate tensile strains of over 50% (*cf.* Supplementary Fig. 3). This mechanical behaviour originates from the anisotropic topology of the 1D rigid $AlO_6$ octahedra chains coupled with flexible hydrogen-bonded interchain contacts. Therefore, although the six MOF configurations exhibit comparable volumetric averages of the elastic modulus (*cf.* Supplementary Table 3), their anisotropic mechanical responses differ significantly.

***Structural transitions and associated local $\mu_2$-OH reorientations between MIL-120(Al) configurations predicted by DFT and MLP calculations.*** Quantitative characterisation of the kinetics of $\mu_2$-OH (re)orientations is essential for determining whether the identified polymorphs correspond to kinetically trapped states or whether they remain dynamically accessible under experimental conditions. For this purpose, we calculated the minimum-energy pathways between the different polymorphic states of MIL-120(Al)s using the climbing image nudged elastic band (CI-NEB) method[19,25] interfaced with the DFT and MLP levels of theory. Representative CI-NEB profiles obtained from DFT (Fig. 3a) and MLP (Fig. 3b) reveal energy barriers of 0.07–0.19 eV per unit cell for the interconversion between the empty configurations. Importantly, the MLP-predicted pathways quantitatively reproduced the DFT-derived barriers and transition-state geometries across all computed systems. The energy profiles and the intermediate images are highly consistent, with almost complete overlap observed between the MLP and DFT curves (*cf.* Supplementary Figs. 4–16 for the complete curves). This quantitative agreement is particularly notable because CI-NEB probes transition-state regions of the potential-energy surface, where interpolation errors often compromise the accuracy of predicted energy barriers. Thus, the MLP's performance confirms that the training set adequately sampled both equilibrium configurations and critical transition geometries.

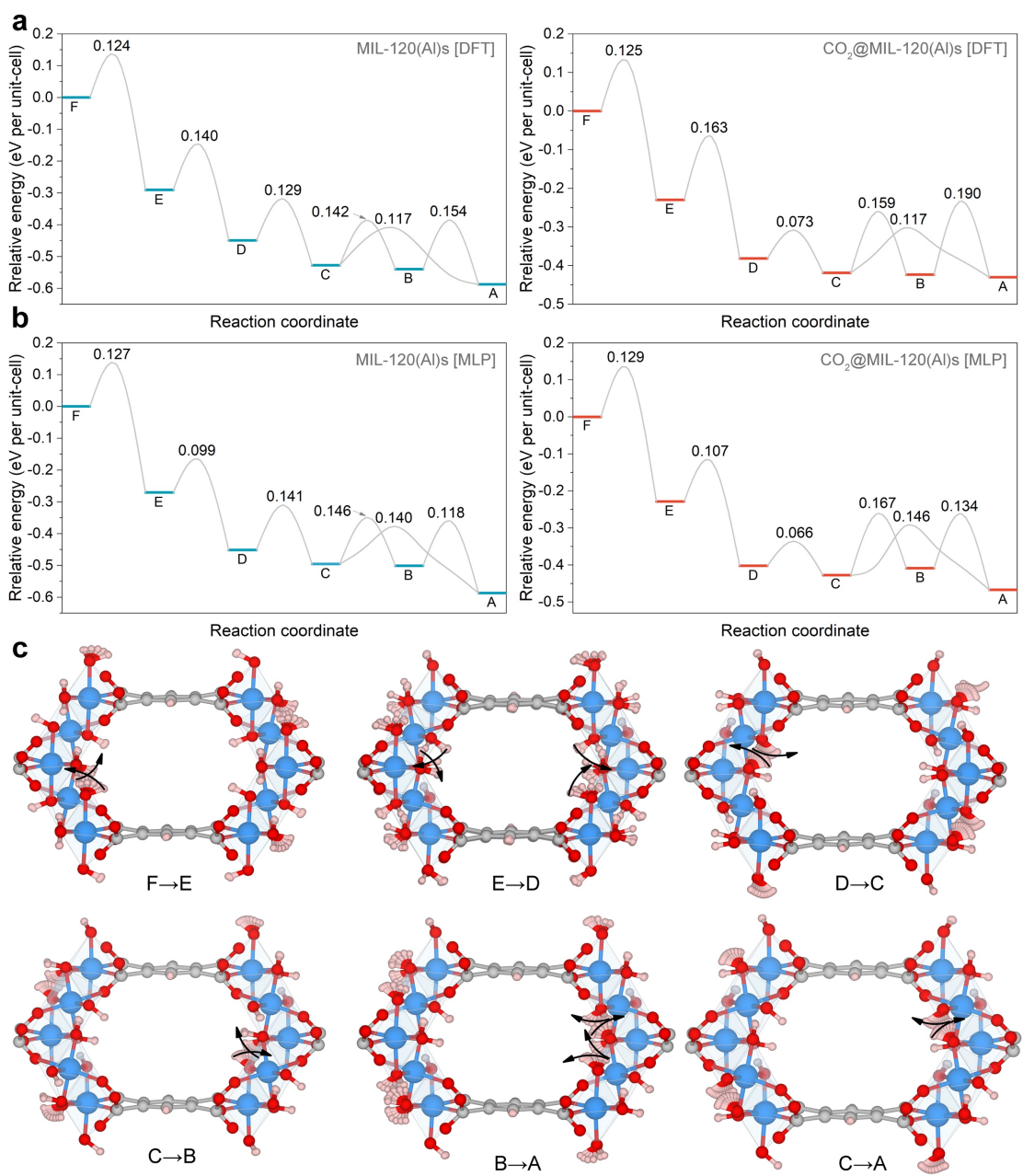

**Fig. 3 | Structural transitions and *μ*₂-OH interconversion pathways of MIL-120(Al)s predicted by DFT- and MLP-CI-NEB calculations**. (**a**) Representative CI-NEB profiles (in eV unit) for the empty (left) and $CO_2$-loaded (right) MIL-120(Al) structures using the DFT-CI-NEB approach. (**b**) Corresponding energy profiles using MLP-driven calculations that closely reproduce DFT barriers. (**c**) Structural snapshots and H atom trajectories illustrating $\mu_2$-OH reorientations for six representative transition paths. The black arrows in the figure indicate the directions in which the $\mu_2$-OH orientation changes. Colour code: Al, blue; O, red; C, grey; H, pink.

Kinetically, the computed interconversion energy barriers (0.07–0.19 eV per unit cell) are readily surmountable at ambient conditions. Interestingly, the presence of $CO_2$ in the MOF pores only slightly affects these energy barriers. For some pathways, $CO_2$ marginally reduces the energy barrier (ΔE↓ ~ 3–8%), indicating that guest-mediated stabilisation of specific transition states occurs through directional host–guest interactions. Analysis of CI-NEB snapshots for the empty MIL-120(Al) structures (Fig. 3c) shows that the structure interconversions imply a dynamic reorientation of the $\mu_2$-OH groups, the H-atom temporarily forming and breaking hydrogen bonds with neighbouring oxygens, primarily with the carboxylate oxygens of the pyromellitate linker, while maintaining its covalent bond with the original $\mu_2$-O. A synchronous H-atom displacement and transient hydrogen-bond rearrangement account for the modest energy barriers involved, highlighting the high degree of freedom of this $\mu_2$-OH group and how it can easily reorient, thereby affecting the features of the accessible MOF pores.

***In-depth microscopic understanding of $CO_2$ adsorption in MIL-120(Al)s by DFT and MLP calculations.*** Next, we assessed how spatial alingments of $\mu_2$-OH in MIL-120(Al)s modulates the adsorption geometries and energetics of $CO_2$ within the MOF channels. To avoid bias towards local minima, the adsorption site locator process employed a random insertion approach, followed by MLP-based relaxation and subsequent DFT refinement of the lowest-energy candidates. The spatial overlap of the $CO_2$ positions optimised using MLP and DFT calculations found to be in excellent agreement. Overlay plots in Fig. 4a show very similar adsorption sites and orientations across all six MOF structure variants, confirming that the MLP captures the local host–guest interactions accurately. In-depth analysis of the orientation of the adsorbed $CO_2$ reveals a clear correlation with the $\mu_2$-OH orientations. In most cases, the $CO_2$ molecular axis is aligned nearly perpendicular to the pore axis. This maximises directional interactions with $\mu_2$-OH groups pointing towards the channel. Both our DFT- and MLP-optimized geometries show that the $O(CO_2)\cdots H(\mu_2$-OH$)$ separating distances are in the range of 2.17~2.74 Å, with both MLP and DFT optimized geometries giving consistent distances and geometries (cf. Fig. 4a). By contrast, in MIL-120(Al)-C and MIL-120(Al)-D, the local $\mu_2$-OH postioning are different: the bridging hydroxyl groups adopt orientations more axial to the channel, which in turn sterically favour a parallel $CO_2$ alignment along the channel. Notably, the presence of $CO_2$ produces only minor adjustments of $\mu_2$-OH orientations relative to the empty frameworks, consistent with the earlier finding that $CO_2$ does not substantially alter reorientation barriers.

In terms of $CO_2$ interaction energies, the MLP's predictions deviate from DFT by at most 2.3 kJ mol$^{-1}$ (*cf.* Fig. 4b). Interestingly, our MLP performs better than other MLPs reported in the literature, such as MACE-MP-0 and fine-tuned DAC-SIM models[26,27] leading to a systematic over- and under-estimation of the interaction energies, respectively (Fig. 4b). This strongly suggests that the transferability of MLPs is limited for MOFs incorporating polar functions and local structural flexibility, herein with the dynamics of the $\mu_2$-OH groups dominating the host–guest interactions. The Widom insertion calculations performed using our trained MLP lead to isosteric heats (Qst) which closely match the experimental values for MIL-120(Al)-A, MIL-120(Al)-B and MIL-120(Al)-C, as shown in Fig. 4c. However, the MIL-120(Al)-F configuration exhibits a computed Qst ~27% higher than the experimental value, while the result based on MACE-DAC is 23% lower. Overall, the stronger host/guest interaction is consistent with the more confined pore environment of the MIL-120(Al)-F characterized by the highly ordered orientation of the $\mu_2$-OH groups towards the pore channels.

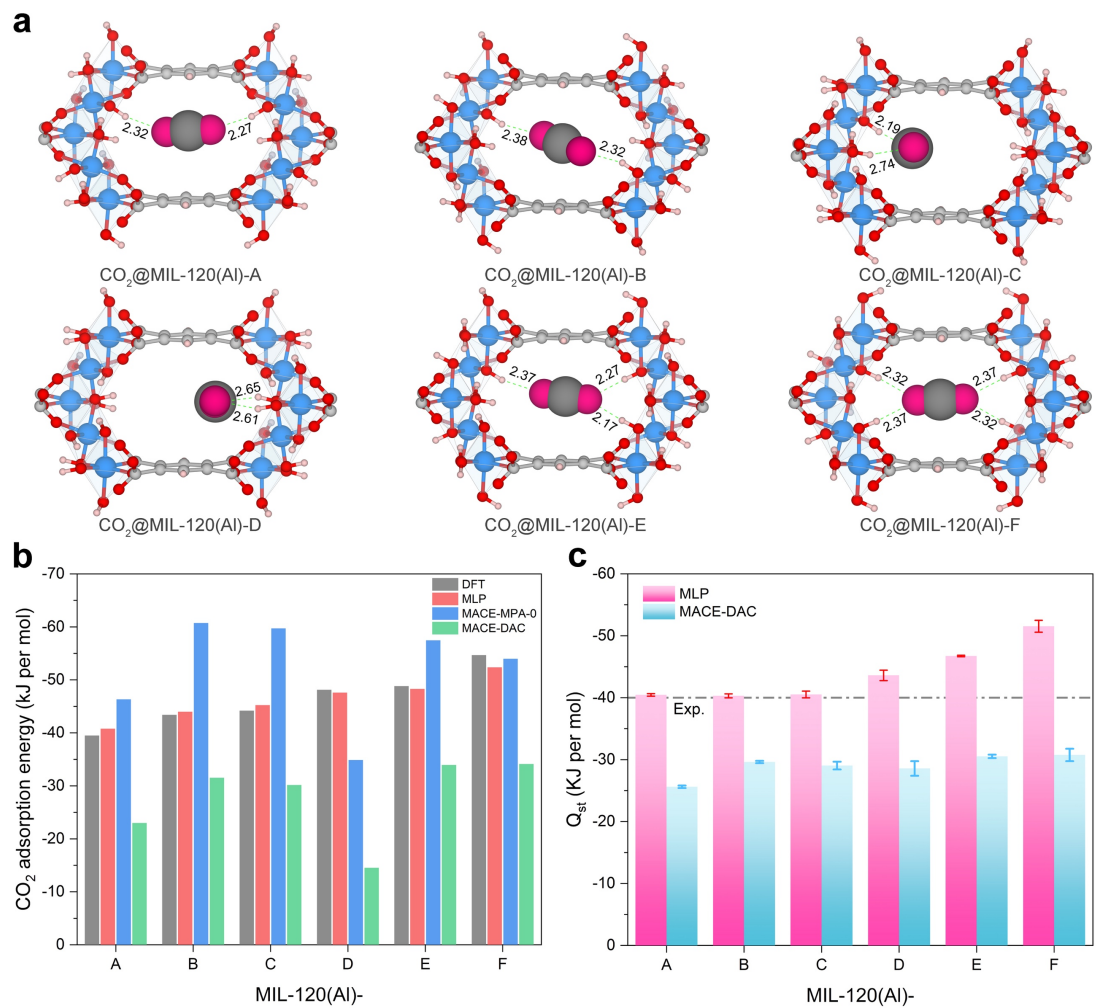

**Fig. 4 | In-depth microscopic understanding of the CO$_2$ adsorption in the diverse MIL-120(Al) configurations**. (**a**) Overlaid lowest-energy CO$_2$ adsorption sites obtained by MLP and DFT calculations for the different MIL-120(Al) structures. The green dashed line highlights the shortest distance between CO$_2$ and $\mu_2$-OH groups in MIL-120(Al) variant structures (unit: Å). It should be noted that the DFT- and MLP-optimised CO$_2$ positions are identical, and therefore indistinguishable. Colour code: Al, blue; O, red; C, grey; H, pink; CO$_2$, magenta and black. (**b**) CO$_2$ interaction energies computed by DFT, our MLP model, and other MLPs including MACE-MPA-0 and MACE-DAC,[26,27] respectively. These interaction energies ($E_{int}$) were computed as $E_{int} = E_{CO_2@MOF} - (E_{MOF} + E_{CO_2})$ using the optimized CO$_2$ molecule, the optimized empty MIL-120(Al)s, and optimized CO$_2$@MIL-120(Al)s at each theoretical level

(MLP/DFT). (**c**) Widom-derived isosteric heats ($Q_{st}$) from our MLP and MACE-DAC). The dashed line represents the experimental $Q_{st}$ value reported earlier.[17]

**Discussions**

In summary, the combination of DFT calculations with a purpose-trained MLP, demonstrates that MIL-120(Al) can adopt a large set of configurations associated with distinct local $\mu_2$-OH orientations. It was found that the commonly adopted structure model in the literature [MIL-120(Al)-F], in which the bridging $\mu_2$-OH point towards pore, is computed to be a high-energy form in the empty scenario, whereas the MIL-120(Al)-A is the more stable configuration with its $\mu_2$-OH groups arranged to form a cooperative, interlocking hydrogen-bond network. Notably, the energy barriers corresponding to the interconversion of these different structure models are relatively small. Phonon calculations showed that all configurations are dynamically stable, and mechanical analysis revealed exceptional ductility along the $AlO_6$ octahedra chain direction. The so-developed MLP trained on an extensive DFT dataset reproduces DFT energetics, CI-NEB energy barriers and phonons for the empty structures, and beyond accurately predicts $CO_2$ adsorption geometries and energies within 2.3 kJ mol$^{-1}$ of DFT values. We found that $CO_2$ does not significantly raise reorientation energy barriers and can even stabilize specific transition states, while conversely $\mu_2$-OH orientation controls whether $CO_2$ binds parallel or perpendicular to the channel and thereby tunes adsorption energetics. Indeed, these computational findings evidence that the local dynamics of the $\mu_2$-OH groups play a major role in the $CO_2$ location and energetics in this ultra-small pore MOF, a feature most often neglected using generic force fields and rigid MOF framework assumption. The precise location of the H atom that cannot be achieved by conventional X-ray diffraction techniques, combined with the flexibility of the $\mu_2$-OH groups, is shown to control the pore aperture size and hence the $CO_2$ adsorption geometries and energetics. More generally, these results also emphasise the necessity of system-specific, high-quality MLP for reliably predicting the adsorption behaviours of guest molecules in ultra-small pore MOFs.

## Methods

### DFT calculations

All DFT calculations were carried out using the Vienna Ab-initio Simulation Package (VASP) code (Version: 5.4.4).[28] The projector augmented wave (PAW) potential and the Perdew-Burke-Ernzerhof (PBE) exchange-correlation functional was adopted.[29,30] An energy cutoff of 650 eV and the Monkhorst-Pack $5 \times 5 \times 6$ k-point grid[31] was chosen to ensure convergence of total energy. These parameters yielded converged total energies and atomic forces within thresholds of $10^{-5}$ eV and $10^{-2}$ eV/Å, respectively. Dispersion interactions were accounted for using the DFT-D3 van der Waals (vdW) correction scheme.[32] Finite-temperature *ab initio* molecular dynamics (AIMD) simulations were carried out in $2 \times 2 \times 2$ supercells to better capture local dynamics and hydrogen-bond rearrangements. AIMD runs used the canonical (NVT) ensemble with a Nosé–Hoover thermostat and a time step of 0.5 fs.[33] The AIMD trajectories were collected at 300, 500 and 800 K for durations exceeding 5 ps. Symmetry constraints were removed in all AIMD simulations to allow unrestricted sampling of local reorientation modes. CI-NEB[19] calculations for transition-state pathways were performed within VASP using standard settings described in the Supplementary Nate2.

### Dataset preparation for MLP training

To build a representative training dataset for the MIL-120(Al)s family, we combined configurations from multiple sampling protocols: DFT geometry optimizations, static single-point calculations, random Widom-style insertions of $CO_2$ at various loadings, structure optimizations under different strains and pressures, CI-NEB intermediate images sampled along $\mu_2$-OH reorientation paths, and AIMD snapshots from both NVT and NPT ensembles. The dataset includes both empty and $CO_2$-loaded MIL-120(Al) configurations in unit-cell and supercell representations. In total 183,061 snapshots were collected. Further details are provided in Supplementary Note 3.

### MLP development

We trained a deep neural network potential using the DeePMD-kit (v2.0.1) implementation of the DeePot-SE framework.[20] The embedding network sizes were set to {25, 50, 100} for successive embedding layers, while the fitting network used three hidden layers of sizes {240, 240, 240}. A radial cutoff of 7.8 Å and a smoothing-length of 1.2 Å were adopted to capture sufficient many-body information while maintaining computational efficiency for the MIL-120(Al)s environments. Training proceeded for 1,000,000 steps with an initial learning rate of $1\times10^{-3}$ that decayed every 5,000 steps; other hyperparameters (batching and loss weightings) follow the protocol summarized in Supplementary Note 4. The choice of cutoff and network depth was validated by convergence tests to ensure robust reproduction of DFT energies and forces for both empty and $CO_2$-loaded configurations.

**MLP-based molecular dynamics and property calculations**

The trained DeePMD potential was also deployed for large-scale molecular dynamics *via* the DeepMD-LAMMPS interface, using the model as a LAMMPS pair style for energy and force evaluations.[34] Phonon calculations were performed using Phonopy and the phonoLAMMPS toolchain on 2×2×2 supercells along a consistent Brillouin zone path.[35,36] CI-NEB pathways were recomputed using the ASE-Python library[37] and the trained MLP to enable extensive transition state sampling at negligible computational cost. The MLP was also used for exhaustive $CO_2$ adsorption site discovery *via* Widom-insertion sampling and adsorption energy scans.

More detailed methodologies are provided in the Supplementary Materials.

## Data availability

All data needed to evaluate the conclusions in the paper are present in the paper and/or the Supplementary Materials. The data related to this article can be accessed online at https://github.com/agrh/Data_Repository_MIL-120_MLP.

## Code availability

The primary packages utilized in this article include VASP (https://www.vasp.at) and DeePMD-kit (https://github.com/deepmodeling/deepmd-kit). Detailed information about the license and the user manual can be found in the abovementioned articles and on their websites.

## Acknowledgments

The computational work was performed using HPC resources from GENCI-CINES (Grant A0180907613). G.M. thanks the Institut Universitaire de France for the Senior Chair.

## Author contributions

D.F. and G.M. designed the research. D.F., F.O. and M.W. carried out the simulations. D.F., F.O., M.W and G.M. wrote the manuscript. G.M. supervised and guided the research.

## Competing interests

The authors declare that they have no competing interests.